\documentstyle[twoside,fancyhea,multicol,pra,aps,epsf,amssymb,jps]{revtex}
\draft
\title{A MARGINAL DIMENSION OF A WEAKLY DILUTED\\ QUENCHED $m$-VECTOR MODEL}

\author{Yu.~Holovatch$^{1,2}$, M.~Dudka$^1$, T.~Yavors'kii$^2$}

\address{$^1$Institute for Condensed Matter Physics of the National Academy
             of Sciences of Ukraine\\
             1 Svientsitskii Str., Lviv, UA--79011, Ukraine\\
             $^2$Ivan Franko National University of Lviv,
             Department for Theoretical Physics\\
             12 Drahomanov Str., Lviv, UA--79005, Ukraine}

\date{Received August 31, 2001}

\begin{document}

\setlength{\jot}{1em}
\setlength{\abovedisplayskip}{1.5em}
\setlength{\belowdisplayskip}{1.5em}
\renewcommand{\baselinestretch}{1}
\renewcommand{\theequation}{\arabic{equation}}

\setcounter{page}{233}

\maketitle

\begin{abstract}
We calculate a marginal order parameter dimension $m_c$
which in a weakly diluted quenched $m$-vector model controls
the crossover from a universality
class of a ``pure'' model ($m>m_c$) to a new universality class ($m<m_c$).
Exploiting the Harris criterion and the field-theoretical renormalization group
approach allows us to obtain $m_c$ as a five-loop $\varepsilon$-expansion as 
well as
a six-loop pseudo-$\varepsilon$ expansion. In order to estimate the numerical
value of $m_c$ we process the series by precisely adjusted
Pad\'e--Borel--Leroy resummation procedures. Our final result
$m_c=1.912\pm0.004<2$ stems from the longer and more reliable
pseudo-$\varepsilon$ expansion, suggesting that a weak quenched disorder
does not change the values of $xy$-model critical exponents as it
follows from the experiments on critical properties of
${\rm He}^4$ in porous media.

\end{abstract}


\medskip

\bgtwocol

\section{Introduction}
\label{I}
Along with the space dimension $d$, the order parameter dimension $m$
is relevant for the universal properties of a model at criticality
\cite{Amit89}. In the
presence of disorder the value of $m$ determines also whether a
disordered model possesses novel universal properties in comparison with a
pure model. A typical example of this feature is presented by
a critical behaviour of a weakly diluted quenched
$m$-vector model \cite{Stanley68}. The model
is defined by a Hamiltonian
\begin{equation}
\label{ham}
H=-\frac{1}{2}\sum_{i,j}J(|{{\bf R}_i}-{{\bf R}_j}|)
\vec{S}_{{\bf R}_i}\vec{S}_{{\bf R}_j}c_{{\bf R}_i}c_{{\bf R}_j},
\end{equation}
where ${\bf R}_i$ span over the sites of a simple cubic lattice and
$\vec{S}_{{\bf R}_i}$ denote the
$m$-component spins interacting via a translationally invariant
short-range isotropic interaction $J(|{\bf R}_i - {\bf R}_j|)$.
The weak disorder is introduced by stochastic uncorrelated
occupation numbers $c_{{\bf R}_i}$ equal to $1$ in the case when
a site is occupied by a magnetic atom and $0$ otherwise
(see Fig.~1). The concentration of occupied sites is
considered to be above the percolation threshold.
The quenched disorder implies
that vacancies $c_{{\bf R}_i}=0$ are fixed and require configurational
averaging of observables \cite{note1}.

The crucial dependence of the
universality class of model (\ref{ham})
on the order parameter dimension $m$ can be
established by the Harris criterion \cite{Harris74}. It states that a
disorder changes the universal critical properties of a ``pure'' model only
if  heat capacity critical exponent of a pure model is positive.
Within the hierarchy of the physical realizations of the $m$-vector model only
the Ising model $m=1$ is characterised by $\alpha=0.109 \pm 0.004 >0 $,
while $xy$- ($m=2$) and Heisenberg ($m=3$) models' heat capacity does not
exhibit divergency at criticality: the corresponding critical exponents
remain negative $\alpha=-0.011 \pm 0.004$, $\alpha=-0.122 \pm 0.009$
\cite{Guida98}. Therefore one can expect that only weakly diluted quenched
Ising model belongs to a new universality class.

Indeed, experimental studies confirm the theoretical prediction. The bulk of
evidence collected in a recent review \cite{Folk01}
demonstrate a novel critical behaviour of the magnetic systems
described by model (\ref{ham}) at $m=1$.
The experimental value of the heat capacity
critical exponent at the $\lambda$-transition in He-4
\cite{Lipa96} corroborated that the system belongs to the $xy$-model
universality class with no divergency of the heat capacity at the phase
transition point.
Subsequently, experiments on the critical behaviour of He-4 in porous media
\cite{He4dil} confirmed the irrelevancy of a weak quenched disorder in this 
case.

\begin{figure}[htbp]
\vspace{4mm}
\label{lattice}
\begin{picture}(60,160)
\setlength{\unitlength}{1mm}
\epsfxsize=50mm
\put(13,3){\epsffile[77 528 337 780]{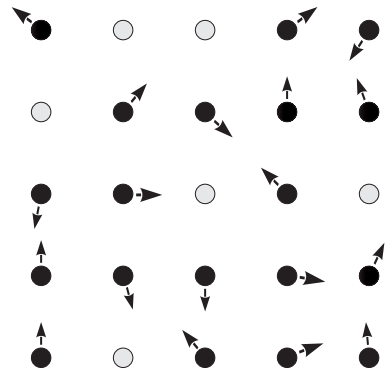}}
\end{picture}
\vspace{5mm}
\caption {The weakly diluted quenched $m$-vector model describes a system of
randomly distributed $m$-component vectors which are fixed on sites of a
three-dimensional cubic lattice and interact via a short-range
translationally-invariant force.}
\end{figure}
\vspace{5mm}

A natural question arises: can  a marginal value of $m$
 be obtained such that for $m>m_c$ the critical exponents of a
weakly diluted quenched $m$-vector model coincide with the corresponding values 
of
a ``pure''model, but for $m<m_c$ they split to new values. Note,
that while the values of the critical exponents both
for pure \cite{Guida98} and diluted $m$-vector \cite{Pelissetto00} models
are calculated precisely, the crossover between two universality classes
has not been the
subject of extended theoretical calculations. One can mention only
the results for $m_c$ on the basis of two- \cite{Holovatch92} and
three-loop renormalization
group studies \cite{Holovatch97} of a weakly diluted $d$-dimensional
$m$-vector model. The obtained values $m_c=2.01$ and $m_c=2.12$ contradict
both experimental data \cite{He4dil} and the Harris criterion \cite{Harris74}.
In contrast, an alternative estimate on the base of a refined by conformal
mapping resummation of the six-loop renormalization group
functions of the $m$-vector model explicitly yielded $m_c=1.942\pm0.026$
\cite{Bervillier86}.
In this study we want to solve the inconsistency of the theoretical results
\cite{Holovatch92,Holovatch97,Bervillier86} and to perform an
independent determination of $m_c$.

Relying on the Harris criterion one can determine $m_c$ from the
requirement of vanishing the heat capacity critical exponent $\alpha$
of the $m$-vector model. We will make use of this condition appealing
to a well-established field-theoretical renormalization group (RG)
approach \cite{Amit89} which allows perturbative calculating of the marginal
order parameter dimension $m_c$. In the next Section we  obtain
$m_c$ as $\varepsilon$- and pseudo-$\varepsilon$ expansions which
follow from the alternative minimal subtraction and massive schemes of
field-theoretical renormalization group approach. In order
to cope with the divergencies of the expansions and
to obtain reliable numerical values  based on them  we  process
the expansions by  appropriate resummation procedures.
The  resummation procedures are also
analyzed and adjusted in the next Section. The outcomes of the study are 
presented
in the concluding Section \ref{III}.

\section{The expansions and the resummation}\label{II}

As  mentioned above,
the marginal dimension $m_c$ of a weakly diluted $m$-vector
model can be reconstituted from the critical properties of a ``pure''
$m$-vector model.
Considering the heat capacity critical exponent $\alpha$ of
a ``pure'' model as a function of the order parameter component number $m$, one
can write the master equation for determining $m_c$ as follows:
\begin{equation}
\label{alpha}
\alpha(m_c)=0.
\end{equation}
The treatment of Eq. (\ref{alpha}) by means of the field-theoretical
RG approach can be performed in various  schemes. Two of them are
most widely used in the theory of critical phenomena. The dimensional
regularisation with the minimal subtraction \cite{tHooft72}
allows obtaining quantities of interest by familiar $\varepsilon$-expansions
\cite{Fisher72} with $\varepsilon=4-d$.
As a starting point to obtain the $\varepsilon$-expansion for $m_c$
serves the
$\varepsilon$-expansion for the critical exponent $\alpha$ of the $m$-vector
model, which is known in the five-loop the approximation \cite{Kleinert91}.
By keeping the coefficients of the expansions as functions of $m$
and reexpanding equation (\ref{alpha}) in $\varepsilon$ we
obtain $m_c$ in the form:
\begin{eqnarray}
\label{epexp}
m_c&=&4- 4\varepsilon+ 4.707199{\varepsilon}^{2}-
8.727517{\varepsilon}^{3} \nonumber\\&&+20.878373{\varepsilon}^{4}.
\end{eqnarray}
Formally, the numerical value of $m_c$  at $d=3$ can be calculated from the
expansion (\ref{epexp}) by the substitution $\varepsilon=1$.

The renormalization conditions of the RG massive scheme \cite{Parisi80}
provide another possibility to obtain the critical properties
 directly at $d=3$. The traditional calculation of the critical
exponents in the massive field theoretical RG scheme implies a numerical
analysis of the RG functions. However, the most accurate estimates of the
critical exponents of a three-dimensional $m$-vector model
\cite{Guida98,Guillou80} are based on a
pseudo-$\varepsilon$ expansion technique
\cite{Nickel}.
This technique avoids intrinsic errors accumulation typical
for the numerical processing of the massive RG functions
and results in a self-consistent collection of
contributions from the different steps of calculations \cite{Guillou80}.
On the
other hand, introducting the auxiliary pseudo-$\varepsilon$ parameter
$\tau$ which mimics the role of the ordinary $\varepsilon$ parameter of the
minimal subtraction RG scheme allows to analyse the series by
methods well-established for the $\varepsilon$-expansion.

In order to obtain a pseudo-$\varepsilon$ expansion for $m_c$ we start from
the expansions of RG functions of $m$-vector model which
within the massive scheme have been calculated in the six-loop approximation
\cite{Sokolov95}.
Introducing into functions \cite{Sokolov95}
the parameter $\tau$ (see \cite{Guillou80} for details)
 we obtain the
pseudo-$\varepsilon$ expansion for $m_c$ at $d=3$ as follows:
\begin{eqnarray}\label{masexp}
m_c&=&4-8/3\tau+ 0.766489{\tau}^{2}-
0.293632{\tau}^{3} \nonumber\\
&&+0.193141{\tau}^{4}-0.192714{\tau}^{5}.
\end{eqnarray}
Again, the resulting numerical value of $m_c$ can be obtained from
expansion (\ref{masexp}) by the final substitution $\tau=1$.

The explicit form of expansions (\ref{epexp}) and (\ref{masexp})
is sufficient in principle to estimate the numerical value of
$m_c$. However, the series for RG functions are known to be of
asymptotic nature \cite{Lipatov77,Guillou77,Brezin78} and
must be resummed before the final substitutions
$\varepsilon=1$ ($\tau=1$). An explicit form of the asymptotic
 of series (\ref{epexp})--(\ref{masexp}) has not
been obtained, contrary to $\varepsilon$-expansions for the $m$-vector model
critical exponents. Consequently, the applicability of resummation procedures to
the series of $m_c$ is only conjectured.

A resummation procedure, which in different modifications is
commonly used in the studies of asymptotic series, is  the
integral Borel transformation \cite{Hardy48}. However, this technique
implies explicit knowledge of the general term of a series and thus cannot
be applied here, where only truncated sums of the series are known.
On the other hand neither can we use the resummation procedures based on
the conformal mapping technique
\cite{Guillou80} since even estimates on large-order order
behaviour of expansions terms are unavailable. In the case of the series
(\ref{epexp})--(\ref{masexp}) we are restricted to the simplest procedures
which do not imply such estimates.

Let us start the analysis by representing series (\ref{epexp})
by means of Pad\'e approximants
$\left[ M/N \right](x)=\sum_{i=0}^M a_i x^i/\sum_{j=0}^N b_j x^j$
in the variable $x=\varepsilon$ \cite{Baker81}.
The result for $m_c$ is shown in the form of a Pad\'e table (\ref{Pademin}),
where the number of the row, $M$, and that of the column, $N$, correspond to the
order of the numerator and that of the denominator of the Pad\'e approximant 
$\left[
M/N\right]$ respectively. Subsequently,
${\it o}$ denotes approximants which can not be constructed
within the considered approximation, while small numbers in (\ref{Pademin})
indicate that approximants have poles (at $\varepsilon=6.52$
and $\varepsilon=1.11$ for $\left[0/2\right]$
and $\left[0/4\right]$ respectively) and thus are unreliable.
\begin{equation}
\label{Pademin}
\left [\begin {array}{ccccc}  4& 2& ^{2.1939}& 1.5086& ^{6.1528}
\\\noalign{\medskip} 0& 2.1624& 2.0316& 1.9365&{\it o}
\\\noalign{\medskip} 4.7072& 1.6493& 1.9208&{\it o}&{\it
o}\\\noalign{\medskip}- 4.0203& 2.1344&{\it o}&{\it
o}&{\it o}\\\noalign{\medskip} 16.858&{\it o}&{\it
o}&{\it o}&{\it o}\end {array}\right ].
\end{equation}
The first column of table (\ref{Pademin})  represents a
straightforward summation of series (\ref{epexp}) terms
and obviously shows the divergence of results with the increase of
the approximation order. Conversely,
the convergence of numbers is observed along the main and next to main
diagonals of the table. For instance, the  expected
inequality $m_c<2$ can be obtained already within the fourth and the fifth
order of the $\varepsilon$-expansion (\ref{epexp}).

The results of Pad\'e-analysis of the pseudo-$\varepsilon$ expansion
(\ref{masexp}) are presented in table (\ref{Pademas}) in the
same notations as in table (\ref{Pademin}).
\begin{equation}
\label{Pademas}
\left [\begin {array}{cccccc}  4& 2.4& 2.0839& 1.9669& 1.9398
& 1.9106\\\noalign{\medskip} 1.3333& 1.9287& ^{1.8799}& 1.9311&
^{2.2425}&{\it o}\\\noalign{\medskip} 2.0998& 1.8875& 1.9084&
 1.9085&{\it o}&{\it o}\\\noalign{\medskip} 1.8062&
 1.9227& 1.9085&{\it o}&{\it o}&{\it o}
\\\noalign{\medskip} 1.9993& 1.9029&{\it o}&{\it o}&{
\it o}&{\it o}\\\noalign{\medskip} 1.8066&{\it o}&{
\it o}&{\it o}&{\it o}&{\it o}\end {array}
\right ].
\end{equation}
Again, small numbers correspond to unreliable
approximants with the poles at $\tau=7.29$ and
$\tau=0.907$ for $\left[1/2\right]$
and $\left[1/4\right]$ respectively.
The distinguishing property of table (\ref{Pademin})
is a convergence of the results on the basis of
a mere summation of the pseudo-$\varepsilon$ expansion terms
(\ref{masexp}) (the first column of table (\ref{Pademas})).
Contrary to the $\varepsilon$-expansion analysis (\ref{Pademin}), the
inequality $m_c<2$ is observed already in a four-loop
approximation and remains valid in the five- and the six-loop approximations.
However, the best convergence is noticed for the approximants parallel
to the main diagonal of table (\ref{Pademas}). For instance,
the six-loop approximants [3/2] and [2/3] as well as the five-loop
approximant [2/2] yield practically the same value
of $m_c$ within the fourth digit leading
to the estimate $m_c=1.9085$. This number is not yet
considered as the most accurate estimate of $m_c$ that one can obtain from
expansion (\ref{masexp}).

By assuming the factorial divergence of the coefficients of expansions
(\ref{epexp}), (\ref{masexp}), one can apply to their analysis a refined
Pad\'e--Borel--Leroy resummation procedure which has been successfully used
in various tasks of theory of critical phenomena \cite{Pade}. The procedure
is based  on the integral Borel transformation
\cite{Hardy48}, however it
uses as an intermediate step an extrapolation by means of a Pad\'e-approximant
\cite{Baker81}.
More precisely, the procedure is defined by the following algorithm:
\begin{itemize}
\item
starting from the initial sum $S$ of $L$ terms one constructs its
Borel--Leroy image
\begin{equation}
\label{BLimage}
S(x)=\sum_{i=0}^La_i x^i\quad\Rightarrow\quad
S^{\rm B}(xt)=\sum_{i=0}^L\frac{a_i(x t)^i}{(i+b)!},
\end{equation}
where $b$ is an arbitrary non-negative number;
\item
subsequently one extrapolates the Borel--Leroy image
(\ref{BLimage}) by a rational
Pad\'e approximant
$$S^{\rm B}(xt)\quad\Rightarrow\quad\left[ M/N \right] (x t);$$
\item
the resummed function $S^{\rm Res}$ is finally  obtained in the following form:
\begin{equation}
\label{res}
S^{\rm Res}(x)=\int_0^\infty\, dt \exp (-t)t^b
\left[ M/N \right] (x t).
\end{equation}
\end{itemize}

\begin{figure}[htbp]
\begin{picture}(80,230)
\setlength{\unitlength}{1mm}
\epsfxsize=85mm
\epsfysize=75mm
\put(-5,4){\epsffile[34 93 531 674]{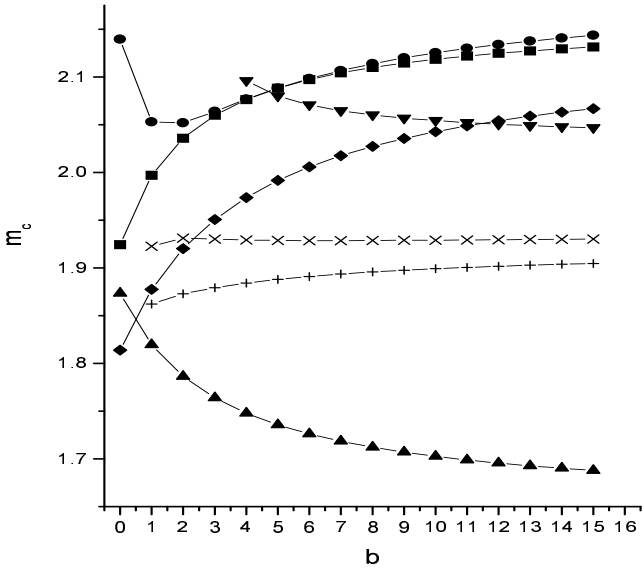}}
\end{picture}
\vspace{5mm}
\caption {\label{eps}
The estimate of the marginal order parameter dimension $m_c$ obtained
on the basis of Pad\'e--Borel--Leroy resummation of the
$\varepsilon$-expansion (\protect{\ref{epexp}}) for
different values of the fit parameter $b$.
The  following symbols show the results
based on different Pad\'e approximants:
 $\blacksquare$ and $\bullet$ mean the [1/1], [0/2]
approximants,
 $\blacktriangle$ and $\blacktriangledown$ show [2/1] and [1/2] approximants,
while  $\blacklozenge$, $+$ and $\times$ denote the [3/1], [2/2] and [1/3]
approximants respectively.
}
\end{figure}

Similar to the Pad\'e-analysis, within the
Pad\'e--Borel--Leroy resummation procedure,
various final estimates of a resummed series can be obtained depending on the 
type of
the Pad\'e approximants chosen. In addition the fit parameter $b$ can be
used for adjusting the resummation procedure to provide a self-consistent 
convergence
of the results. In Fig.~\ref{eps} we present the result of the
$\varepsilon$-expansion processing by means of the resummation
procedure (\ref{BLimage})--(\ref{res}).
Here, the estimates of $m_c$ on the basis of the higher three-, four- and
five-loop
$\varepsilon$-expansion are depicted depending on
the approximant type as well as on the fit parameter $b$.
Among all the possibilities we chose the approximants
close to the main diagonal, namely [1/1], [0/2] in three-,
[2/1], [1/2] in four and [3/1], [2/2], [1/3] in the five-loop approximation.
Notably, approximants having poles  on the real positive semiaxis  are
considered as unreliable.
For instance, Fig.~\ref{eps} displays the real part of values
with imaginary parts smaller than $10^{-7}$.

Since no information is available on the large-order behaviour
of series (\ref{epexp}),
all the values of $b$ can be considered as equally suitable.
An evident property of all the curves in Fig.~\ref{eps}
is a saturation of the value of $m_c$ for large values of $b$.
In order to exclude the values on the basis of large $b$ and taking into
account that in similar tasks of critical phenomena $b$ accounts for several
units, we restrict ourselves only to the values of $b$ within the interval
$0\leq b\leq 5$. We neither take into account the approximant
$[1/2]$ since it provides only two values within the interval.

\vspace{5mm}
\vspace{7mm}

To obtain a confidence interval for $m_c$ we
analyse the data of Fig.~\ref{eps} in the following manner.
As a first step we obtain the central value of $m_c$ on falling look on the data 
on
each approximant separately.
To this end we average all the values on the basis of each approximant over
$b$, $0\leq b\leq 5$. The corresponding error bars are then determined
by a half of the difference between the maximal and minimal values of $m_c$ 
within the
considered interval of $b$. The resulting estimates of $m_c$
in the successive number of loops are presented in Fig.~\ref{er_eps}, a,
where the result on the basis of each working approximant is shown by its
central value (black circles) and the error bars.
For instance, in the three-loop approximation two confidence intervals
correspond to [1/1] (a smaller central value) and [0/2] (a larger central
value) approximants; the four-loop approximation contributes by a
single estimate on the basis of [2/1] approximant, while the five-loop
approximation yields three confidence intervals from the [1/3] (the
largest central value), [3/1] (the middle central value) and [2/2]
approximants.

In order to get a final estimate of $m_c$ within each order of
perturbation theory we take up the values on the basis of different
approximants of the same order as independent.
Doing so allows us to consider their average as
an overall estimate of $m_c$ within a given number of loops.
The overall estimates of $m_c$ obtained in this way
depend on the loop order oscillatively, which permits to suppose that
a successive central value lies between two preceding central values.
Thus we average the overall estimate on the basis of the
approximation order with a corresponding overall estimate from
the preceding order, choosing the error bars as a half of a difference
between maximal and minimal values from these pairs.
Finally, we obtain a sequence of confidence intervals of $m_c$
such that error bars of a higher result lie
completely within error bars of the previous one.

Based on the analysis described above
we obtain the following estimates of $m_c$ from the
$\varepsilon$-expansion (\ref{epexp}):
\begin{eqnarray}
\nonumber&{\rm 4LA:}\quad m_c=1.996\pm0.104,&\\
&{\rm 5LA:}\quad m_c=1.923\pm0.051.&\label{epsres}
\end{eqnarray}

One can note from estimates (\ref{epsres}) that only the five-loop
$\varepsilon$-expansion excludes the values $m_c\ge2$, and therefore
brings about a weakly quenched disorder irrelevancy for the
$xy$-model universality class.

\begin{figure}
\vspace{2mm}
\begin{picture}(80,230)
\setlength{\unitlength}{1mm}
\epsfxsize=85mm
\epsfysize=75mm
\put(-5,4){\epsffile[28 97 531 679]{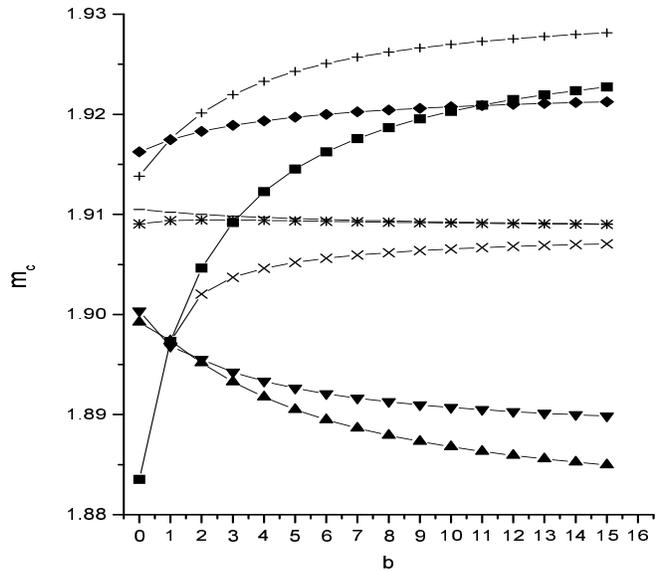}}
\end{picture}
\vspace{2mm}
\caption{\label{mas}
The estimate of the marginal order parameter dimension $m_c$ obtained
on the basis of Pad\'e--Borel--Leroy resummation of the
pseudo-$\varepsilon$ expansion (\protect{\ref{masexp}}) for
different values of the fit parameter $b$.
The following symbols show the results
based on different Pad\'e approximants:
$\blacksquare$, $\blacktriangle$, $\blacktriangledown$ denote the results
on the base of [1/1], [2/1], [1/2] approximants,
$\blacklozenge$, $+$ and $\times$ show the [3/1], [1/3], [2/2] approximants,
while  $\ast$ and $-$ correspond to [2/3] and [3/2] approximants
respectively.
}
\vspace{2mm}
\end{figure}

Let us turn now to the Pad\'e--Borel--Leroy analysis of the
pseudo-$\varepsilon$ expansion (\protect{\ref{masexp}}).
The dependencies of the estimates of $m_c$ on the basis of the
pseudo-$\varepsilon$ expansion
on the fit parameter $b$ are shown for different approximants types
increasing number of loops from three to six
in Fig.~\ref{mas}.
Since the pseudo-$\varepsilon$ expansion is one order longer,
in addition to the approximants of Fig.~\ref{eps}
we use here the  approximants of the
six-loop order, namely
the near-diagonal approximants [3/2] and [2/3]. The values
of $m_c$ obtained on the basis of the approximant
[0/2] do not fit Fig.~\ref{mas} and thus are not presented,
though they will be taken into account in the final calculations.

Applying the procedure identical to the $\varepsilon$-expansion
we obtain the resulting estimates of $m_c$ depending on the
type of the approximant as presented in Fig.~\ref{er_eps}, b.
For instance, in the three-loop approximation two confidence intervals
correspond to [1/1] (smaller central value) and [0/2] (larger central value)
approximants; the four-loop approximation contributes by
two almost identical estimates on the basis of [1/2]
and [2/1] approximants, while the five-loop
and the six-loop approximations yield
three and two confidence intervals respectively.
Subsequently, averaging the overall estimate on the basis of a given
loop order with the corresponding overall estimate from
a preceding order, one obtains a sequence of convergent confidence intervals
for $m_c$:
\begin{eqnarray}\nonumber
&{\rm 4LA:}\quad m_c=1.976\pm0.117,&\\
\label{psepsres} \nonumber
&{\rm 5LA:}\quad m_c=1.904\pm0.013,\\
&{\rm 6LA:}\quad m_c=1.912\pm0.004.&
\end{eqnarray}
The sequence of results (\ref{psepsres}) obtained on the basis of the
pseudo-$\varepsilon$
expansion (\ref{masexp})  shows evidently better convergence properties than the
corresponding results (\ref{epsres}) obtained on the basis of the
$\varepsilon$-expansion (\ref{epexp}). For instance, the error bar of the
five-loop $\varepsilon$-expansion estimate is four times larger than the
corresponding estimate on the basis of pseudo-$\varepsilon$.

\edtwocol
\vspace{5mm}
\begin{figure}[htbp]
\begin{picture}(180,200)
\epsfxsize=65mm
\epsfysize=70mm
\put(20,4){\epsffile[42 91 501 734]{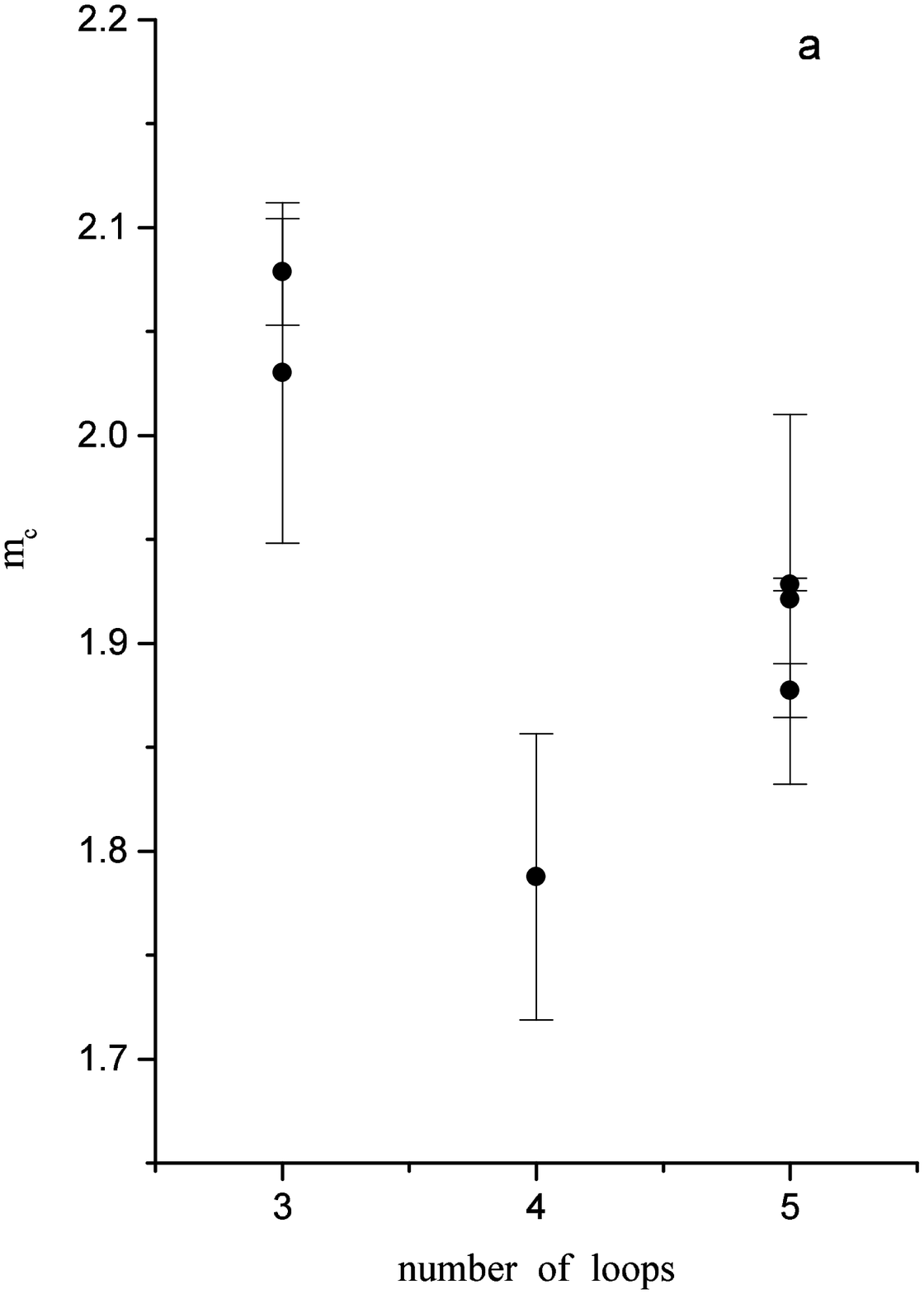}}
\epsfxsize=65mm
\epsfysize=70mm
\put(270,4){\epsffile[81 96 501 729]{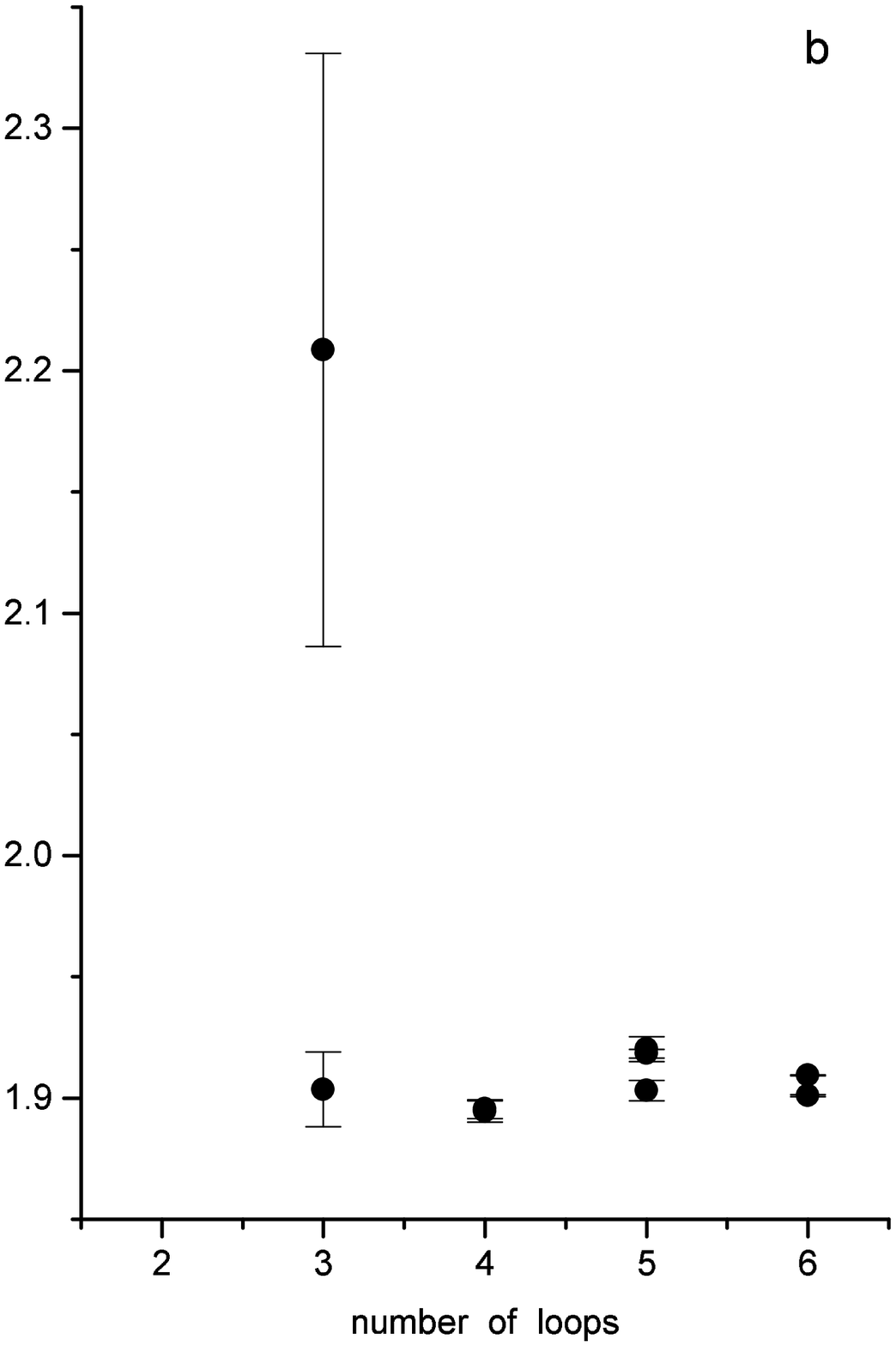}}
\end{picture}
\vspace{9mm}
\caption{\label{er_eps}
The confidence intervals for the marginal order parameter dimension $m_c$
of a weakly diluted $m$-vector model as functions of the order of approximation.
The left picture presents the result of the Pad\'e--Borel--Leroy resummation of
the $\varepsilon$-expansion (\protect{\ref{epexp}}), the right-hand picture
corresponds to the analysis of the pseudo-$\varepsilon$ expansion 
(\protect{\ref{masexp}}).
See the text for the whole description.}
\end{figure}

\vspace{3mm}

\bgtwocol
\section{Conclusions}
\label{III}
In the theory of critical phenomena the values of critical exponents
and critical amplitudes ratios have attracted considerable interest of 
researchers.
This study, however, has calculated another important quantity, which
is responsible for a crossover phenomenon in the weakly diluted quenched
$m$-vector model. Based on the Harris criterion as well as on the
field-theoretical renormalization group approach we
determined the marginal order parameter dimension $m_c$ of the model.
Before our work only one reliable numerical estimate
of $m_c$ was obtained \cite{Bervillier86}. The result $m_c=1.942\pm0.026$
\cite{Bervillier86} stemmed from the application of a conformal mapping
resummation technique to the massive six-loop RG functions at $d=3$.
We completed this study with two more results from two different RG schemes.
Within the minimal subtraction
scheme $m_c$ is obtained as an $\varepsilon$-expansion
(\ref{epexp}), while within
the massive scheme $m_c$ is calculated as a pseudo-$\varepsilon$
expansion
(\ref{masexp}). Both series allow to estimate a numerical value
of $m_c$ after their analysis by Pad\'e--Borel--Leroy
resummation procedure, however the final estimate of the paper
$m_c=1.912\pm0.004$ is based on the pseudo-$\varepsilon$
expansion. This is explained not only by a longer series
calculated for the expansion, but also by its
better convergent properties in comparison with the $\varepsilon$-expansion.
Such a situation has been already observed for a calculation of a
marginal dimension of a cubic model \cite{Folk00b}.

Though the obtained value of $m_c$ is very close to the integer value 2,
our estimate evidently shows that within the error bars
it is slightly smaller than 2.
This result implies that universal properties of the
$xy$-model are not affected by a weakly quenched disorder at criticality.
This conclusion is confirmed by recent experimental studies on the
critical behaviour of the superfluid He$^4$ in porous media \cite{He4dil}.

We thank Claude Bervillier for useful discussions  and Ted Knoy for a
 thorough reading of the manuscript.
\edtwocol

\bigskip

\begin{referen}

\bibitem{Amit89}
See, e.~g. D.~J.~Amit {\it Field Theory, the Renormalization Group, and Critical
Phenomena} (World Scientist, Singapore, 1989).

\bibitem{Stanley68}
H.~E.~Stanley,
Phys. Rev. Lett {\bf 20}, 589 (1968).

\bibitem{note1}
R.~Brout, Phys. Rev. {\bf 115}, 824 (1959).

\bibitem{Harris74}
A.~B.~Harris, J. Phys. C {\bf 7}, 1671 (1974).

\bibitem{Guida98}
R.~Guida, J.~Zinn-Justin,
J. Phys. A {\bf 31}, 8103 (1998).

\bibitem{Folk01}
R.~Folk, Yu.~Holovatch, T.~Yavors'kii, preprint
cond-mat/0106468 (2001), to appear
in: Physics Uspekhi (2002).

\bibitem{Lipa96}
The original result $\alpha=-0.01285 \pm 0.00038$ presented in Ref.
[J.~A.~Lipa, D.~R.~Swanson, J.~A.~Nissen,
T.~C.~P.~Chui, U.~E.~Israelsson, Phys. Rev.  Lett. {\bf 76}, 944 (1996)]
is incorrect. However, a new estimate reported in Ref.
[J.~A.~Lipa {\em et al.}, Phys. Rev. Lett. {\bf 84}, 4894, (2000)]
also yields a negative value of the exponent: $\alpha=-0.01056\pm0.00038$.

\bibitem{He4dil}
J.~Yoon, M.~H.~W.~Chan,
Phys. Rev. Lett. {\bf 78}, 4801 (1997);
G.~M.~Zassenhaus, J.~D.~Reppy, Phys. Rev. Lett. {\bf 83}, 4800 (1999).

\bibitem{Pelissetto00}
A.~Pelissetto, E.~Vicari,
Phys. Rev. B {\bf 62} 6393 (2000).

\bibitem{Holovatch92}
J. Jug, Phys. Rev B {\bf 27}, 609 (1983);
Yu.~Holovatch, M.~Shpot, J. Stat. Phys. {\bf 66}, 867 (1992).

\bibitem{Holovatch97}
Yu.~Holovatch, T.~Yavors'kii,
Cond. Mat. Phys. {\bf 11}, 87 (1997);
Yu.~Holovatch, T.~Yavors'kii, J. Stat. Phys. {\bf 92}, 785 (1998).

\bibitem{Bervillier86}
C.~Bervillier, Phys. Rev. B {\bf 34}, 8141 (1986).

\bibitem{tHooft72}
G.'t Hooft, M.~Veltman,
Nucl. Phys. B {\bf 44}, 189 (1972);
G.'t Hooft, Nucl. Phys. B {\bf 61}, 455 (1973).

\bibitem{Fisher72}
K.~G.~Wilson, M.~E.~Fisher,
Phys. Rev. Lett. {\bf 28}, 240 (1972).

\bibitem{Kleinert91}
H.~Kleinert, J.~Neu, V.~Schul\-te-Froh\-lin\-de, K.~G.~Che\-tyr\-kin,
S.~A.~Larin,
Phys. Lett. B {\bf 272}, 39 (1991); Erratum:
Phys. Lett. B {\bf 319}, 545 (1993).

\bibitem{Parisi80}
G.~Parisi (1973), unpublished;
G.~Parisi, J. Stat. Phys. {\bf 23}, 49 (1980).

\bibitem{Guillou80}
J.~C.~Le Guillou, J.~Zinn-Justin, Phys. Rev. B {\bf 21}, 3976 (1980).

\bibitem{Nickel}
The pseudo-$\varepsilon$ expansion was introduced by B.~G.~Nickel,
see citation 19 in Ref. \cite{Guillou80}.

\bibitem{Sokolov95}
S.~A.~Antonenko, A.~I.~Sokolov,
Phys. Rev. E {\bf 51}, 1894 (1995).

\bibitem{Lipatov77}
L.~Lipatov,
Sov. Phys. JETP {\bf 45}, 216 (1977).

\bibitem{Guillou77}
E.~Br\'ezin, J.~Le Guillou, J.~Zinn-Justin,
Phys. Rev. D {\bf 15}, 1544 (1977).

\bibitem{Brezin78}
E.~Br\'ezin, G.~Parisi,
J. Stat. Phys. {\bf 19}, 269 (1978).

\bibitem{Hardy48}
G.~H.~Hardy,
{\it Divergent Series} (Oxford, 1948).

\bibitem{Baker81}
G.~A.~Baker, Jn, P.~Graves-Morris,
{\it Pad\'e Approximants} (Addison-Wesley: Reading, MA, 1981).

\bibitem{Pade}
G.~A.~Baker, B.~G.~Nickel, M.~S.~Green, D.~I.~Meiron,
Phys. Rev. Lett. {\bf 36}, 1351 (1976);
G.~A.~Baker, B.~G.~Nickel, D.~I.~Meiron,
Phys. Rev. B {\bf 17} 1365 (1978).

\bibitem{Folk00b}
R.~Folk, Yu.~Holovatch, T.~Yavors'kii Phys. Rev. B {\bf 62}, 12 195
(2000); Erratum: Phys. Rev. B {\bf 63}, 189 901 (2001)
\end{referen}
\end{document}